\documentclass[11pt,a4paper]{article}
\pdfoutput=1 

\usepackage[utf8x]{inputenc}
\usepackage{epic}
\usepackage{graphicx}

\usepackage[T1]{fontenc}
\usepackage{amssymb,amsmath,amsfonts}
\usepackage{stackrel}
\usepackage{float}
\usepackage{dsfont}
\usepackage{jheppub} 

\newcommand{\mn}{{\mu \nu}}
\newcommand{\coup}{g_{\rm YM}}

\newcommand{\be}{\begin{equation}}
\newcommand{\ee}{\end{equation}}
\newcommand{\bse}{\begin{subequations}}
\newcommand{\ese}{\end{subequations}}
\newcommand{\ba}{\begin{eqnarray}}
\newcommand{\ea}{\end{eqnarray}}
\newcommand{\bea}{\begin{eqnarray}}
\newcommand{\eea}{\end{eqnarray}}

\newcommand{\TT}{\mathcal{T}}
\newcommand{\HT}{\mathcal{H}}
\newcommand{\AT}{\mathcal{A}}
\newcommand{\gT}{g}
\newcommand{\g}{g}
\newcommand{\gB}{g}
\newcommand{\h}{h}
\newcommand{\dl}{\chi}
\newcommand{\tint}{t_{\mathrm{int}}}
\newcommand{\tYM}{t_{\mathrm{YM}}}

\begin{document}

\title{
Time evolution of a toy semiholographic glasma}

\author[a]{Christian Ecker,}
\author[a,b]{Ayan Mukhopadhyay,}
\author[a]{Florian Preis,}
\author[a]{Anton Rebhan}
\author[a]{and Alexander Soloviev}

\affiliation[a]{Institut f\"{u}r Theoretische Physik, Technische Universit\"{a}t Wien,\\
Wiedner Hauptstr.~8-10, A-1040 Vienna, Austria}

\affiliation[b]{Department of Physics, Indian Institute of Technology Madras,\\ Chennai 600036, India}

\emailAdd{christian.ecker@tuwien.ac.at}
\emailAdd{ayan@physics.iitm.ac.in}
\emailAdd{fpreis@hep.itp.tuwien.ac.at}
\emailAdd{rebhana@hep.itp.tuwien.ac.at}
\emailAdd{alexander.soloviev@tuwien.ac.at}

\abstract{We extend our previous study of a toy model for coupling classical Yang-Mills equations for describing
overoccupied gluons at the saturation scale with a strongly coupled infrared
sector modeled by AdS/CFT. Including propagating modes in the bulk we find that the Yang-Mills sector loses its initial energy to a growing black hole
in the gravity dual such that there is a conserved energy-momentum tensor
for the total system while entropy grows monotonically. This involves a numerical AdS simulation with a backreacted boundary source
far from equilibrium.
} 

\maketitle

\section{Introduction}
In the ultrarelativistic heavy-ion collider experiments at RHIC and the LHC, a strongly interacting 
deconfined state of matter,
the so-called quark-gluon plasma (QGP), is produced.
On extremely short time-scales, there appears to arise hydrodynamic behavior 
with a very low value of the specific viscosity \cite{Heinz:2013th,Romatschke:2017ejr}, making the QGP arguably the most
perfect fluid ever observed and presenting a challenge to a fundamental description
based on perturbative QCD, which should govern the high-energy partonic degrees of freedom
of the colliding nuclei
with relatively small running coupling $\alpha_s=g^2/4\pi$.

In the color-glass-condensate (CGC) framework \cite{Gelis:2010nm}, strong interactions
are due to the large occupation numbers of order $1/\alpha_s$ of the gluons liberated in the early
stages of the collision, which appear at a semi-hard momentum scale $Q_s$, the so-called saturation
scale, with $Q_s\gg\Lambda_{\rm QCD}$ at RHIC and LHC.  
This out-of-equilibrium gluonic matter with high occupation numbers, dubbed glasma \cite{Lappi:2006fp}, 
can be approximated by numerical solutions of classical Yang-Mills equations \cite{Romatschke:2006nk,Gelis:2013rba,Berges:2013fga}
until occupation numbers become much smaller than $1/\alpha_s$, where a kinetic description
may take over \cite{Kurkela:2015qoa}.
However, this glasma copiously emits relatively soft gluons
with momenta below the scale $Q_s$ where the running coupling is much stronger,
producing a medium of soft gluons
which is expected to play a dominant role in bottom-up thermalization according to \cite{Baier:2000sb,Kurkela:2011ub}.

In gauge/gravity dual descriptions of heavy-ion collisions \cite{DeWolfe:2013cua,Chesler:2013lia,Chesler:2015lsa}, the perturbative QCD part of this evolution
is usually ignored 
or used only to provide initial conditions for a subsequent holographic treatment, with no backreaction
of the strongly coupled soft sector of the system on the more weakly coupled hard sector
\cite{vanderSchee:2013pia,Casalderrey-Solana:2014bpa}.
Such a backreaction appears to be particularly interesting given the fact that
the gravity-dual description leads to
the qualitatively different top-down thermalization scenario where high-energy modes are the first to attain
a thermal distribution \cite{Balasubramanian:2011ur}.\footnote{A certain modification towards bottom-up thermalization
behavior has been found when including higher-derivative corrections to 
Einstein gravity \cite{Steineder:2012si, Stricker:2013lma,Waeber:2018bea}, which are however known only
as corrections in lowest nontrivial power of the inverse 't Hooft coupling.}

In \cite{Iancu:2014ava}, a `semi-holographic' attempt to combine a description of the hard degrees of freedom
in terms of glasma field equations
and of soft strongly self interacting degrees of freedom in terms of gauge/gravity duality has been presented,
where each sector modifies the dynamics of the complement sector through a minimal local coupling of
the respective dimension-four operators. In \cite{Mukhopadhyay:2015smb} this approach was extended and slightly corrected 
to permit the construction of a conserved energy-momentum tensor of the total system, which is living in
physical Minkowski space, while the gravity-dual sector has a nontrivial boundary metric. (As discussed in
\cite{Banerjee:2017ozx,Kurkela:2018dku} it is also possible, and ultimately desirable, to have dynamics of the weakly
coupled sector take place in yet another nontrivial auxiliary metric.)
Moreover, Ref.~\cite{Mukhopadhyay:2015smb} has set up an extremely simple toy model for studying the combined
system numerically. This was done by restricting the couplings to one between the two energy-momentum tensors
pertaining to the Yang-Mills sector and the holographic sector, but in a spatially homogeneous and isotropic situation
where the gravity side did not allow for propagating degrees of freedom. Nevertheless, nontrivial dynamics
was observed that consisted of oscillatory energy transfer between the Yang-Mills system and the holographic side,
with conservation of the total energy, but no thermalization. While the holographic side included a seed black hole,
the latter could not grow absent any propagating modes in the bulk. The toy model of Ref.~\cite{Mukhopadhyay:2015smb}
instead permitted controlled studies by having a closed-form solution on the gravitational side and thus
a proof-of-concept of the proposed iterative procedure, which indeed showed quick convergence.

In order to study thermalization, which requires propagating degrees of freedom in the bulk that can
build up a black hole, one has to either relax the restriction to spatial homogeneity and isotropy or
include a coupling of scalar operators as indeed initially proposed. Computationally this means
having to deal with numerical AdS calculations with a nontrivial boundary source that gets updated
in subsequent iterations of solving the complete system. In this paper we present the successful
implementation of this scheme in a variant of the toy model introduced in Ref.~\cite{Mukhopadhyay:2015smb},
but reduced by one spatial dimension such that one deals with AdS$_4$ instead of AdS$_5$ geometries
where it is easier to achieve high numerical accuracy.
We expect, however, that this simplification already shows the main features present in the AdS$_5$/SYM$_4$ system
to be explored in future work.

\section{The Setup}
In accordance with the CGC and glasma description of the early stages of the QGP, we model the hard degrees of freedom, i.e. dynamics at the saturation scale $Q_s$, by a classical Yang-Mills theory \cite{Gelis:2010nm,Lappi:2006fp}. Following the semiholographic approach to the QGP proposed in \cite{Iancu:2014ava, Mukhopadhyay:2015smb} we extend the glasma picture by including nonperturbative effects of soft degrees of freedom utilizing the gauge-gravity duality. In doing so we replace non-perturbative QCD by $\mathcal{N}=4$ super Yang-Mills theory which at infinite coupling and a large number of colors allows for a dual description in terms of classical supergravity (SUGRA).

The interaction between the hard and soft sectors is established by promoting the marginal couplings of each sector to functions of gauge-invariant operators of their complements. The (coarse-grained) operators at our disposal in the effective description of the soft sector are the energy momentum tensor $\TT^\mn$, the glueball density operator $\HT$ and the Pontryagin density $\AT$. For the scope of this paper we will restrict ourselves to the coupling of the scalar operator $\HT$ only. This is obtained from the generating functional by varying with respect to the source, e.g.
\begin{equation}
\HT=\frac{1}{\sqrt{\gT}}\frac{\delta W[\h]}{\delta \h},\label{eq:ScalarEVfromW}
\end{equation}
where $W$, $\gT_\mn$, and $\h$ denote the generating functional, the metric of the space-time, and the source, respectively. The background metric $\gT_\mn$ in our context only serves as a computational device and will be set to the Minkowski metric $\eta_\mn$ eventually.  Note that a non-trivial choice for the source corresponds to a marginal deformation of the theory.

In order to simplify the simulation presented in the following sections we will model the hard sector by a classical Yang-Mills theory in $2+1$ dimensions and the soft sector by a gravitational theory in $3+1$ dimensions. Nevertheless, we are confident that the qualitative behavior of the system will not change in a $3+1$ dimensional space-time. In the remainder of this section the number of space-time dimensions $d$ will be kept arbitrary.

Starting from the classical Yang-Mills action we can easily deform the hard sector with a scalar operator by adding a source term
\begin{equation}
S_\mathrm{YM}=-\frac{1}{4\coup^2}\int\mathrm{d}^dx\sqrt{-\g}\left(1+\dl(x)\right)F_\mn^a F^{a\mu\nu},
\end{equation}
with the Yang-Mills coupling constant $\coup$. The non-Abelian field strength in terms of the gauge field is given by $F_{\mu\nu}=\partial_\mu A_\nu-\partial_\nu A_\mu-i [A_\mu,A_\nu]$.
First of all we notice that physically this deformation amounts to locally rescaling the Yang-Mills coupling constant $\coup$. Second, a calculation of the reponse analogous to \eqref{eq:ScalarEVfromW}
yields
\begin{equation}
\frac{1}{\sqrt{\gT}}\frac{\delta S_\mathrm{YM}}{\delta \dl}=-\frac{1}{4\coup^2} F_\mn^a F^{a\mu\nu}.
\end{equation}

In a next step we bring the two deformed sectors into contact by simply adding $S_\mathrm{YM}[A_\mu,\dl]$ and $W[\h]$ supplemented with an interaction term for the two scalar deformations\footnote{A similar action capturing the effective metric couplings has been discussed in Appendix A.2 in \cite{Kurkela:2018dku}.}
\begin{equation}
S = -\frac{1}{4\coup^2}\int\mathrm{d}^dx\sqrt{-\g}\left(1+\dl(x)\right)F_\mn^a F^{a\mu\nu} + W[\h] - \frac{Q_s^d}{\beta}\int\mathrm{d}^dx\sqrt{-\gB}\h\dl.\label{eq:fullaction}
\end{equation}
The saturation scale $Q_s$ appears for dimensional reasons and is accompanied by a phenomenological dimensionless free parameter $\beta$ which allows for tuning the interaction of both sectors.

Inspecting \eqref{eq:fullaction} immediately reveals that the two scalar fields, $\HT$ and $\h$, are auxiliary fields. Their equations of motion, given by
\begin{equation}
\h=-\frac{\beta}{4Q_s^d\coup^2} F_\mn^a F^{a\mu\nu}, \quad\quad \dl = \frac{\beta}{Q_s^d}\HT,\label{eq:auxeom}
\end{equation}
indeed connect the deformations of each sector to gauge independent single trace operators of the respective other sector.
After integrating out $\HT$ and $\h$ the action becomes
\begin{equation}
S=-\frac{1}{4\coup^2}\int\mathrm{d}^dx\sqrt{-\g}F_\mn^a F^{a\mu\nu} + W\left[-\frac{\beta}{4Q_s^d\coup^2} F_\mn^a F^{a\mu\nu}\right].\label{eq:fullaction2}
\end{equation}
This form of the action is discussed in general including the tensor and pseudoscalar coupling channels\footnote{The special case considered here is obtained from the formulation in  \cite{Mukhopadhyay:2015smb} by omitting the couplings $\gamma$ and $\alpha$.} in \cite{Mukhopadhyay:2015smb}. In a recent publication \cite{Kurkela:2018dku} the tensor channel (which is omitted here) was further improved upon by employing a so-called democratic coupling approach.

Let us now turn to the equations of motion for the gauge field $A_\mu$ and the calculation of the expectation value $\HT$. The former read
\begin{equation}
D_\mu\left[\left(1+\frac{\beta}{Q_s^d}\HT\right)F^{a\mn}\right]=0,\label{eq:defYMequation}
\end{equation}
where we introduced the gauge covariant derivative $D_\mu=\nabla_\mu - i  A^a_\mu T^a$ with $\nabla_\mu$ denoting the Levi-Civita connection of the background metric $\g_\mn$.
For calculating $\HT$ we employ the holographic dictionary, which maps the generating functional $W$ to the $(d+1)$ dimensional SUGRA (on-shell) action and operators to fields in the gravity theory satisfying asymptotically anti-de Sitter (AdS) boundary conditions. The relevant terms of the action for our purposes are given by the Einstein--Hilbert action with a cosmological constant coupled to a massless Klein-Gordon scalar field 
\begin{equation}\label{grav_action}
S_{hol}= \frac{1}{2\kappa}\int \text{d}^{d+1} x\sqrt{-G}\left(R-2\Lambda-\frac{1}{2}(\nabla \phi)^2\right)\,,
\end{equation}
where the cosmological constant is $\Lambda=-\frac{(d-1)(d-2)}{2L^2}$ with the AdS radius $L$ set to unity in the remainder of the paper.
The bulk equations of motion arising from \eqref{grav_action} are 
\begin{align}\label{grav_eom}
G^{MN}\nabla_M\nabla_N\phi&=0\nonumber\,,\\
R_{MN}-\frac{1}{2}R G_{MN}-3 G_{MN}&=\kappa(\nabla_M \phi\nabla_N\phi-\frac{1}{2}G_{MN}(\nabla \phi)^2) \, .
\end{align}

In Fefferman-Graham coordinates, where $\rho=0$ denotes the location of the conformal boundary of the (d+1) dimensional space-time, the metric and the scalar field have the following asymptotic expansions \cite{deHaro:2000vlm}
\begin{align}
G_\mn&=\frac{1}{\rho}\left(g_{(0)\mu\nu}+\ldots+\rho^{d/2}g_{(d)\mu\nu}+\mathcal{O}(\rho^{d/2} \log (\rho))\right)\label{eq:G_expansion},\\
\phi&=\phi_{(0)}+\ldots +\rho^{d/2}\phi_{(d)}+\mathcal{O}(\rho^{d/2} \log (\rho))\label{eq:phi_expansion},
\end{align}
from which we can read off the expectation values $\TT_\mn=\frac{d}{\kappa}g_{(d)\mu\nu}+X_{\mu\nu}$ and $\HT=\frac{d}{\kappa}\phi_{(d)} + \psi_{(d)}$, where $X_{\mu\nu}$ and 
$\psi_{(d)}$ are local functionals of the boundary sources. The leading coefficient in the metric expansion is fixed by the background metric $g_{(0)\mu\nu}=\gB_\mn$. The non-normalizeable mode of the scalar field $\phi_{(0)}$ is dual to the source in the generating functional $W$, which in our setup is related to the Lagrange density of the classical Yang-Mills sector
\begin{equation}
\phi_{(0)}=-\frac{\beta}{4Q_s^d\coup^2} F_\mn^a F^{a\mu\nu}\,.
\end{equation}

The total energy momentum tensor of our semiholographic model is
\begin{eqnarray}
T^\mn &=& \tYM^\mn + \TT^\mn + \tint^\mn\nonumber\\
&=&  \frac{1}{\coup^2}\left(1+\frac{\beta}{Q_s^d}\HT\right)\left(F^{\mu\alpha}F^\nu_{\phantom{\nu}\alpha}-\frac{1}{4}\gB^\mn F_{\alpha\beta}F^{\alpha\beta}\right) + \TT^\mn -\h\HT\gB^\mn\,.\label{Eq:EMT_total}
\end{eqnarray}
Each of the three contributions is obtained by the variation of the corresponding term in \eqref{eq:fullaction} with respect to $\gB_\mn$ and employing Eqs. \eqref{eq:auxeom}. We will refer to the first contribution as the Yang-Mills energy momentum tensor, the second as the holographic energy momentum tensor and to the third as the interaction energy momentum tensor. Note that the expression here differs from the corresponding one in \cite{Mukhopadhyay:2015smb} where the terms in $\tYM^\mn$ proportional to $\HT$ were assigned to the interaction energy momentum tensor. Although the assignment of the contributions to the different sectors is somewhat arbitrary, the advantage of the form presented here is that $\tint^\mn$ only consists of the agents responsible for the deformation, in our case the two simplest gauge independent single trace scalar operators.

The conservation of the energy momentum tensor $\nabla_\mu T^\mn=0$ is implied by separate Ward identities in the respective sectors of our model
\begin{eqnarray}
\nabla_\mu \TT^\mn&=&\HT\partial^\nu\h,\label{eq:Ward_hol}\\
\nabla_\mu \tYM^\mn&=&\frac{Q_s^d}{\beta}\h\partial^\nu\left(1+\frac{\beta}{Q_s^d}\HT\right).\label{eq:Ward_YM}
\end{eqnarray}
The sum of the terms on the right hand side is precisely $-\nabla_\mu \tint^\mn$. Furthermore, we also want to mention the trace anomalies of the individual sectors, which read
\begin{eqnarray}
\gB_\mn\TT^\mn&=&(d-4)\HT \h+A,\label{eq:trace_hol}\\
\gB_\mn\tYM^\mn&=&(d-4)\left(1+\frac{\beta}{Q_s^d}\HT\right)\frac{Q_s^d}{\beta}\h,\label{eq:trace_YM}
\end{eqnarray}
where $A$ denotes the holographic conformal anomaly, which is a local functional of the boundary sources and vanishes for the case considered below. Note that in general even if both $\TT^\mn$
and $\tYM^\mn$ are tracefree, eg. for $\gB_\mn=\eta_\mn$ and $d=4$, the full system is not conformal due to the contribution $\gB_\mn \tint^\mn=-d\ \HT\h$.

\subsection{Classical Yang-Mills sector}

We will now work in a $d=2+1$ dimensional space-time with $\gB_\mn = \eta_\mn$ and restrict to isotropic homogeneous SU(2) color gauge fields in temporal gauge, $A^a_0=0, A^3_0=0$ with $a=1,2$. To further simplify this toy model as far as possible, we make $\tYM^\mn$ diagonal by assuming color-space locking, $A^a_i = \delta^a_i f(t)$ and $A^3_i=0$ with $i=1,2$. 
The single remaining degree of freedom $f(t)$ satisfies an equation for an anharmonic oscillator with time dependent damping obtained from \eqref{eq:defYMequation}\footnote{Eq.~\eqref{YM_f} is an unforced damped nonlinear Duffing equation. The latter appears in many contexts and has been extensively studied  \cite{duff,BRAVOYUSTE1986347}.}
\begin{equation}\label{YM_f}
f^{\prime \prime}(t)+f(t)^3=f^\prime(t)\frac{\beta \mathcal{H}^\prime}{1+\frac{\beta}{Q_s^3} \mathcal{H}}\,.
\end{equation}
The energy density and the pressure are
\begin{equation}\label{Eq:YM_EMT}
\varepsilon=\frac{1+\frac{\beta}{Q_s^3} \mathcal{H}}{2\coup^2 }(2f^\prime(t)^2+f(t)^4),\quad p= \frac{1+\frac{\beta}{Q_s^3} \mathcal{H}}{2\coup^2 } f(t)^4\,,
\end{equation}
The source for the dilaton in terms of the YM fields is given by \eqref{eq:auxeom} 
\begin{align}\label{source}
\h&=\frac{\beta}{2Q_s^3\coup^2}(2f^{\prime2}-f^4).
\end{align}

\subsection{Holographic sector}

To be consistent with the YM sector we also impose homogeneity and isotropy in the spatial field theory directions of the bulk theory of the holographic sector. We make the following ans{\"a}tze for the metric and the massless scalar field in in-going Eddington-Finkelstein coordinates
\begin{align}\label{Eq:bulkMetric}
ds^2=-A(r,v)dv^2+2dvdr+S^2(r,v)(dx^2_1+dx^2_2),\quad \phi=\phi(r,v)\,.
\end{align}

The equations of motion \eqref{grav_eom} then take the following form\footnote{It is interesting to note that these equations are equivalent to those of a homogeneous but anisotropic black brane without scalar matter.}
\begin{align}
S''&=-\frac{\kappa}{4} S \left(\phi'\right)^2\,,\label{Eq:charEOM}\\
\dot{S}'&=\frac{3 S}{2}-\frac{\dot{S}S'}{S}\,,\label{Eq:charEOM2}\\
\dot{\phi}'&=-\frac{\dot{S} \phi'}{S}-\frac{ \dot{\phi} S'}{S}\,,\label{Eq:charEOM3}\\
A''&=\frac{4 \dot{S} S'}{S^2}-\kappa\dot{\phi}\phi'\,,\label{Eq:charEOM4}\\
\ddot{S}&=\frac{\dot{S} A'}{2}-\frac{\kappa\dot{\phi}^2 S}{4}\,,\label{constraint}
\end{align}
where prime denotes radial derivatives $f^\prime=\partial_r f$ and the dot-derivative is defined as $\dot{f}=\partial_v f  -\frac{1}{2}A(r,v)\partial_r f$.
Near the boundary ($r=\infty$) solutions to these equations can be expressed as power series in $r$
\begin{align}
A(r,v)&=r^2 \sum_{n=0}^{\infty}a_{n}(v) r^{-n}\,, \\
S(r,v)&=r\sum_{n=0}^{\infty}s_{n}(v) r^{-n}\,,\\
\phi(r,v)&=\kappa\sum_{n=0}^{\infty}\phi_{n}(v)r^{-n}\,.
\end{align}
Fixing the conformal boundary metric to Minkowski $ds_{b}^2=r^2\eta_{\mu\nu}dx^\mu dx^\nu$ determines the leading coefficients $a_{0}=1$ and $s_{0}=1$, and the residual gauge freedom $r \rightarrow r + \xi(v)$ is fixed by setting the subleading coefficient $a_{1}=0$.
Solving the equations order by order in $r$ gives
\begin{align}
A(r,v)&=r^2-\frac{3}{4}\phi_0'(v)^2+ a_3(v)\frac{1}{r}+\mathcal{O}(r^{-2})\label{eq:AEFseries}\,,\\
S(r,v)&=r-\frac{1}{8} \phi_0'(v)^2\frac{1}{r} +\frac{1}{384} \left(\phi_0'(v)^4-48 \phi_3(v) \phi_0'(v)\right)\frac{1}{r^3}+\mathcal{O}(r^{-4})\label{eq:SEFseries}\,,\\
\phi(r,v)&=\phi_0(v)+\phi_0'(v)\frac{1}{r}+\phi_3(v)\frac{1}{r^3}+\mathcal{O}(r^{-4})\label{eq:phiEFseries}\,,
\end{align}
where the normalizable modes $\phi_3(v)$ and $a_3(v)$ remain undetermined in this procedure and need to be extracted from the full bulk solution.
Furthermore one obtains the relation
\begin{align}
a_3'(v)&=\frac{1}{8} \left(12 \phi_3\phi_0'(v)-3 \phi_0'(v)^4+4 \phi_0'''(v) \phi_0'(v)\right)\,.\label{eq:Ward_ser}
\end{align} 

In order to identify the expectation values of the energy momentum tensor and the scalar operator it is convenient to asymptotically transform the series solutions \eqref{eq:AEFseries} and \eqref{eq:phiEFseries} to Fefferman-Graham coordinates \eqref{eq:G_expansion}.
The relevant coefficients in the Fefferman-Graham expansion in terms of their Eddington-Finkelstein counterparts are given by
\begin{align}
\phi_{(0)}=\phi_0,\quad \phi_{(3)}=\phi_{3}+\frac{1}{3} \phi_{0}'''-\frac{1}{4}(\phi_0')^3, \quad g_{(3)ij}=\frac{1}{3}\mathrm{diag}(-2 a_{3},a_{3},a_{3}) \,.
\end{align}
The expectation values of the energy momentum tensor and the scalar operator are then given by
\begin{align}
\mathcal{T}_{\mu\nu}&=\frac{3}{\kappa}g_{(3)\mu\nu}=\frac{1}{\kappa}\mathrm{diag}(-2a_{3},-a_{3},-a_{3})\,,\label{Eq:EMT_hol}\\
\langle \mathcal{O}\rangle&\equiv \mathcal{H} =\frac{3}{\kappa}\phi_{(3)}=\frac{3}{\kappa}(\phi_{3}+\frac{1}{3} \phi_{0}'''-\frac{1}{4}(\phi_0')^3)\,.\label{Eq:VEV_hol}
\end{align}
Evaluating the holographic Ward identity \eqref{eq:Ward_hol} reproduces the relation \eqref{eq:Ward_ser} we find from solving the near boundary expansion 
\begin{equation}\label{Eq:Ward_hol}
({g_{(3)0}}^0)'=\phi_{(3)}\phi_{(0)}\,.
\end{equation}


\section{Solution procedure}
In this section we describe how we obtain solutions for the time evolution problem of the coupled system \eqref{YM_f}, \eqref{Eq:charEOM}--\eqref{constraint} for given values for the couplings $\beta$ and $\coup$ and given initial energies in the Yang-Mills sector ($\epsilon_{YM}^{ini}$) and the holographic sector ($\epsilon_{hol}^{ini}$).

Our approach is to solve the coupled system with an iterative procedure that starts with an initial guess for $f(t)$ which we get from a solution to \eqref{YM_f} for $\beta=0$ 
\begin{equation}\label{Eq:freeYM}
f(t)^{\prime\prime}+2f(t)^3=0\,.
\end{equation}
Solutions to \eqref{Eq:freeYM} can be written in terms of the Jacobi elliptic function
\begin{equation}\label{Eq:f_ini}
f(t)=\pm \sqrt[4]{2C}\, \text{sn}\left(\left.\sqrt[4]{\frac{C}{2}}(t-t_0)\right| -1\right)\,,
\end{equation}
where the integration constant $C/\coup^2=\epsilon_{YM}^{ini}$ can be identified via \eqref{Eq:YM_EMT} with the initial energy in the YM-sector. Without loss of generality we set $t_0=0$, which corresponds to our initial time.
From the initial guess \eqref{Eq:f_ini} and a chosen value for $\beta$ we compute via \eqref{source} the time dependent boundary source for the gravity system $\phi_{(0)}(t)=h(t)$.

For our model defined by \eqref{eq:fullaction2} it is sufficient to consider only positive values of the coupling $\beta$. Switching the sign of $\beta$ amounts to switching the sign of $h(t)$ which we identify with the boundary source $\phi_{(0)}$ of the scalar field in the gravitational bulk. Absent any (odd) scalar self interactions the action given in \eqref{grav_action} is invariant under the transformation $\phi \to -\phi$. Therefore, the holographic contribution to the energy momentum tensor will not change, while the expectation value $\HT$ changes sign. However, in the Yang-Mills equation \eqref{eq:defYMequation} as well as in the Yang-Mills energy momentum tensor only the combination $\beta\HT$ appears, which are therefore invariant under $\beta\to -\beta$. Finally, the product $h\HT$ in the exchange part of the full energy momentum tensor is likewise invariant.

We numerically evolve the boundary sourced gravity system with the spectral method explained in \cite{Chesler:2013lia}, using 20 Chebyshev grid points in the holographic direction and a $4^{th}$ order Adams-Bashforth time stepping algorithm with step size $\Delta t=1/800$.
In order to get a well defined initial value problem resulting in a stable time evolution it is necessary to choose a computational domain in the bulk direction that contains the apparent horizon $r_{ah}$, defined by $\dot{S}(t,r)|_{r_{ah}}=0$, on the initial slice.\footnote{Note that other authors \cite{Chesler:2010bi,Casalderrey-Solana:2013aba,Attems:2016tby} use the residual gauge freedom $r\rightarrow r+\xi(v)$ to fix the apparent horizon to a constant value in the radial direction which is then used to bound the computational domain.} 
Initial data for the gravity system are fixed by $a_{3}(t=0)=-{\epsilon_{hol}^{ini}}/{2}$ and a radial profile for the scalar field which evaluates to $\phi(r,t=0)=-\beta \epsilon_{YM}^{ini}$ for the initial guess \eqref{Eq:f_ini} in combination with the Ward identy \eqref{Eq:Ward_hol}. To measure the accuracy of our numerical scheme we monitor in each time step the violation of the constraint equation \eqref{constraint} and the Ward identity \eqref{Eq:Ward_hol} whose absolute values we demand to be smaller than $10^{-6}$. 
From the solution of the gravity problem we extract, via \eqref{Eq:EMT_hol} and \eqref{Eq:VEV_hol}, the time evolution of $\mathcal{T}_{\mu\nu}(t)$ and $\mathcal{H}(t)$ respectively.
As a next step we numerically solve the Yang-Mills equation \eqref{YM_f} for the new $f(t)$ with the Mathematica routine NDSolve, using $f'(0)$ and $f(0)$ from \eqref{Eq:f_ini} as initial conditions and the result for $\mathcal{H}(t)$ from the gravity simulation. 
This completes the first iteration.

With the new $f(t)$ we can evaluate the total energy momentum tensor \eqref{Eq:EMT_total} and check if it is conserved in time. This will typically not be the case after the first iteration and we have to iterate again. To initialize the next iteration we have to compute the new source for the gravity simulation. Compared to the initial iteration, where we provide an analytic guess for $f(t)$, the updated $f(t)$ in subsequent iterations is known only numerically which introduces via \eqref{source} some numerical noise in the new source for the gravity simulation. We reduce the numerical noise in the source with a low-pass filter before we feed it to the gravity code. As filtering tool we use the Mathematica routine LowpassFilter and choose a cutoff frequency of 0.1 and filter kernel of length 1.
We continue this iterative procedure until the total energy is conserved to $\mathcal{O}(10^{-5})$ or better.
The iterative procedure is summarized in Fig. \ref{loop}. In Appendix \ref{app:1} we discuss the procedure in more detail.
\begin{figure}[!ht]
\begin{center}
\includegraphics[width=0.9\textwidth]{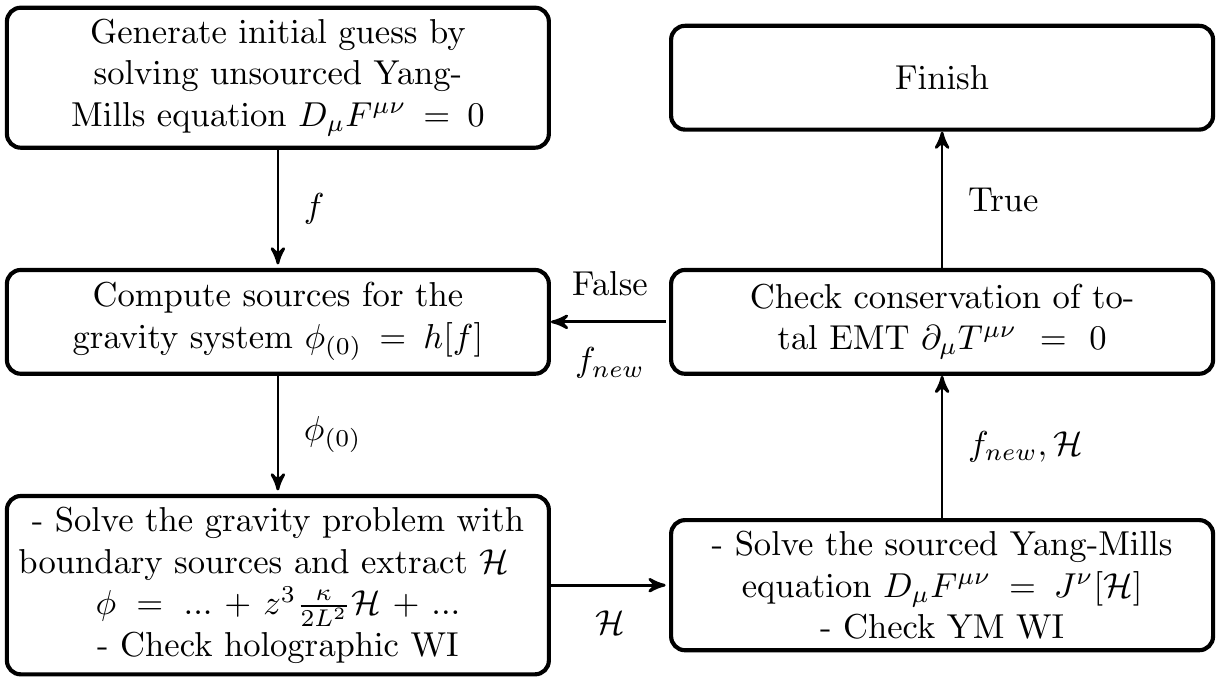}
\caption{Flow chart of the iterative procedure explained in the main text.}
\label{loop}
\end{center}
\end{figure}

\section{Results}
\begin{figure}[!ht]
        \centering
\begin{minipage}{0.5\textwidth}
        \centering
        \includegraphics[width=\textwidth]{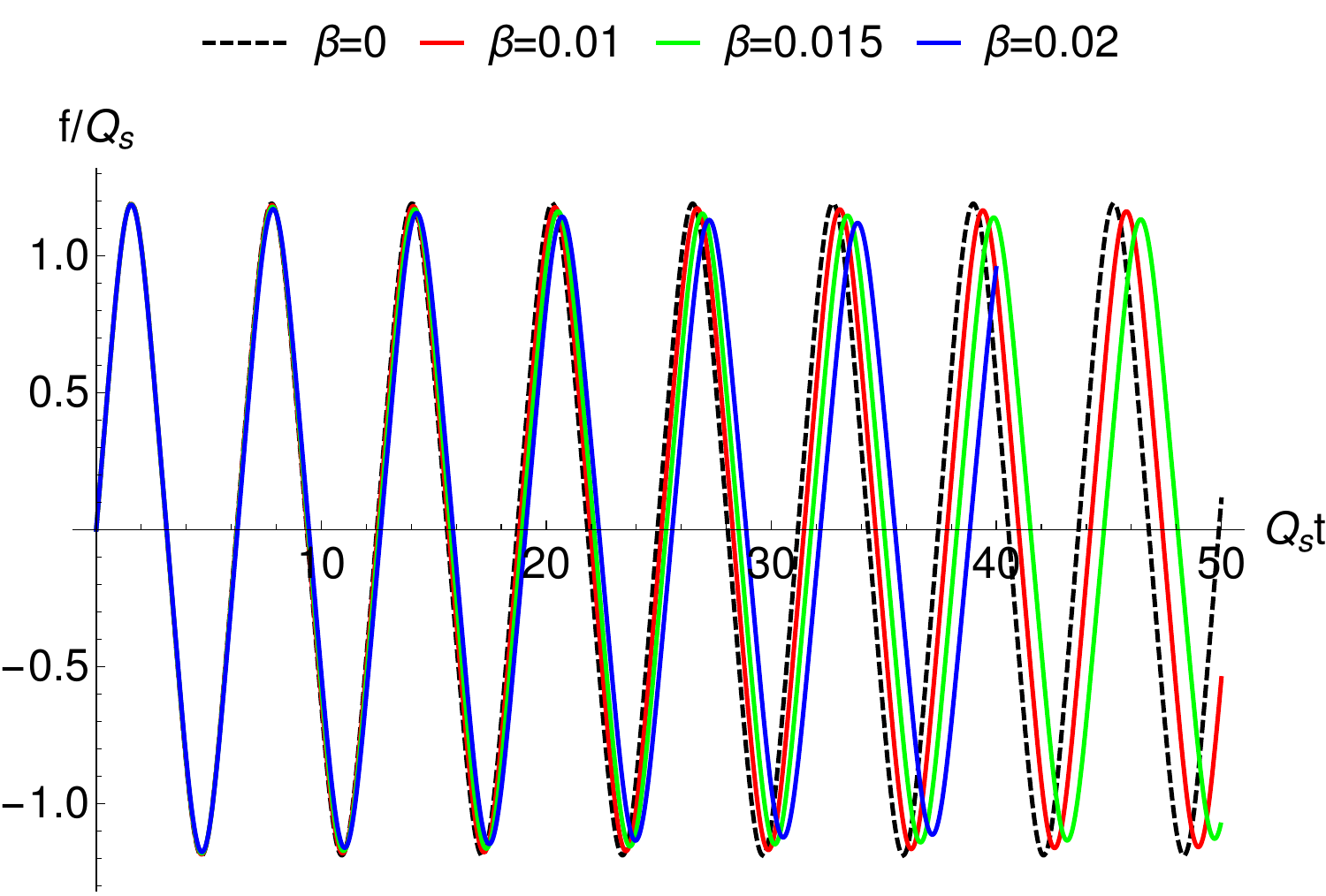}
    \end{minipage}\hfill
    \begin{minipage}{0.5\textwidth}
        \centering
    \includegraphics[width=\textwidth]{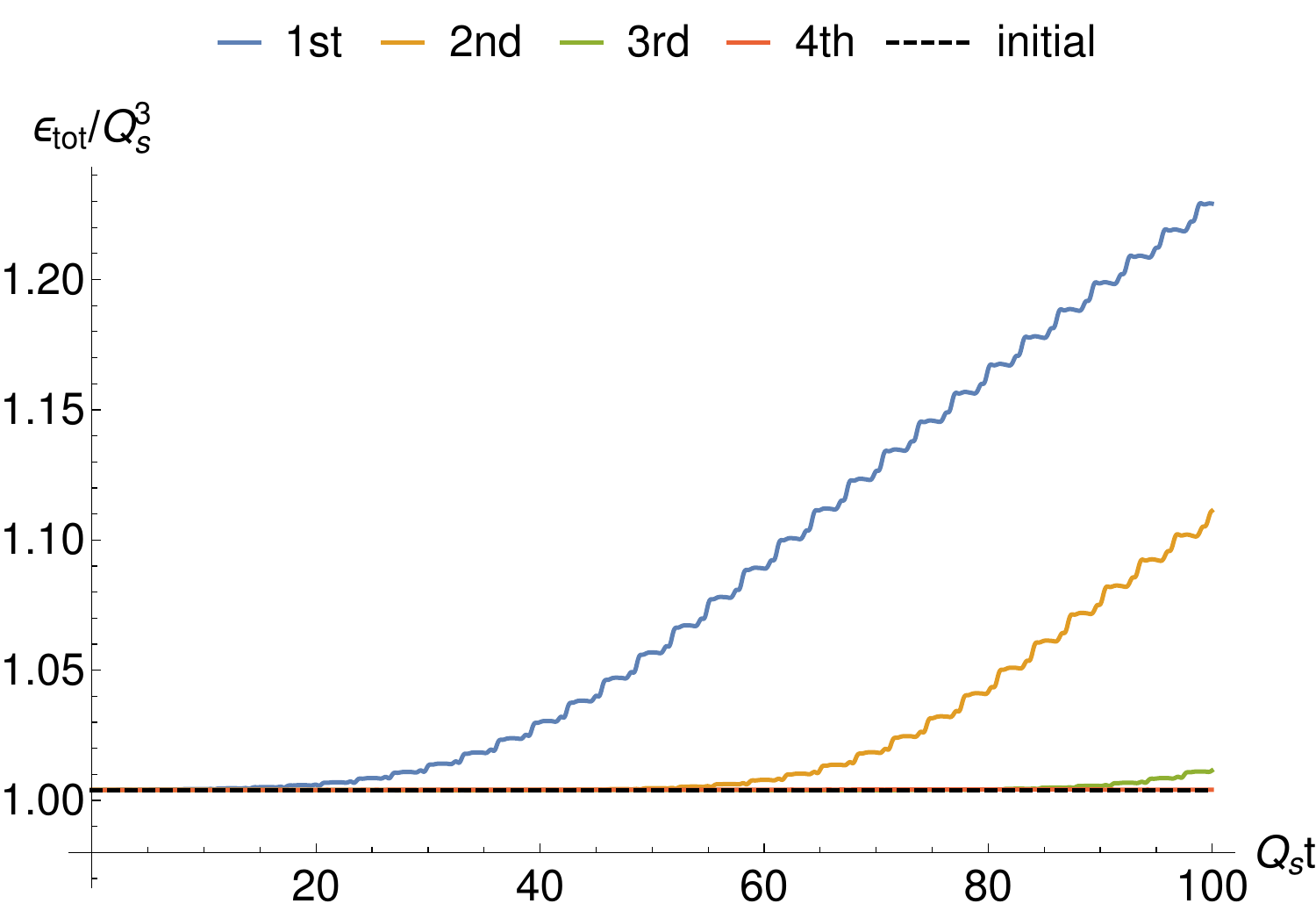}
    \end{minipage}\hfill
\caption{Left:  The YM gauge field. Right: Total energy for $\beta=0.01$ as function of time for subsequent iterations.}
\label{f_vev}
\end{figure}

Following the procedure outlined above, we compute the gauge field degree of freedom $f(t)$, displayed in the left panel of Fig. \ref{f_vev} 
for three different values of $\beta$ after four iterations. For the other parameters we chose the initial Yang-Mills energy density to be $\epsilon_\mathrm{YM}/Q_s^3=1$, the initial energy of the strongly coupled sector to be $\epsilon_\mathrm{hol}/Q_s^3=0.004$ and the Yang-Mills coupling as $\coup/\sqrt{Q_s}=1$. As mentioned above, the criterion for an acceptable solution is to yield a constant total energy of the system. The right panel in Fig. \ref{f_vev} clearly shows that with each iteration the total energy for the choice $\beta=0.01$ stays on its initial value for a longer period of time.

\begin{figure}[!ht]
    \centering
    \begin{minipage}{0.55\textwidth}
        \includegraphics[width=\textwidth]{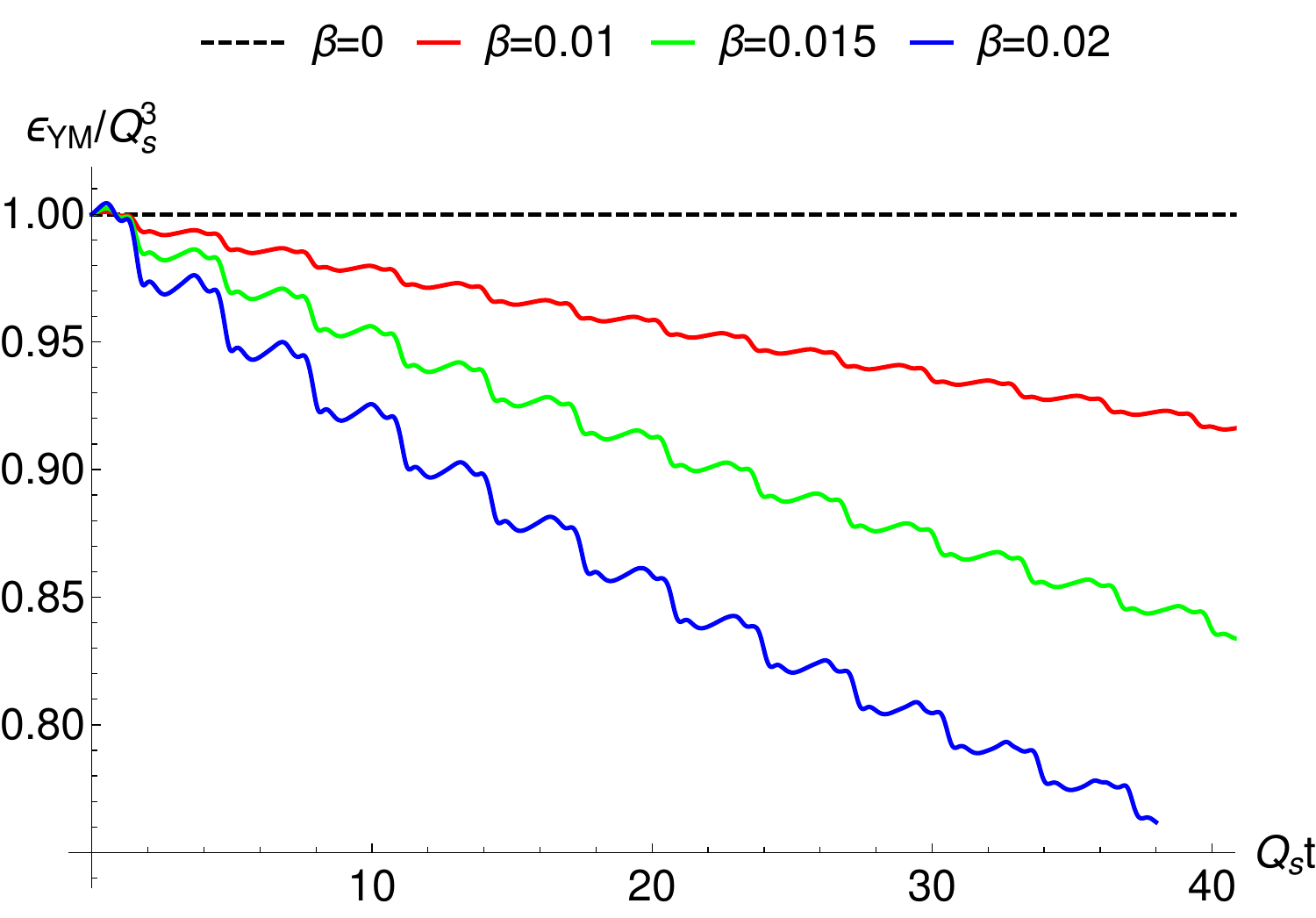}
            \end{minipage}\vfill
         \begin{minipage}{0.55\textwidth}
         \includegraphics[width=\textwidth]{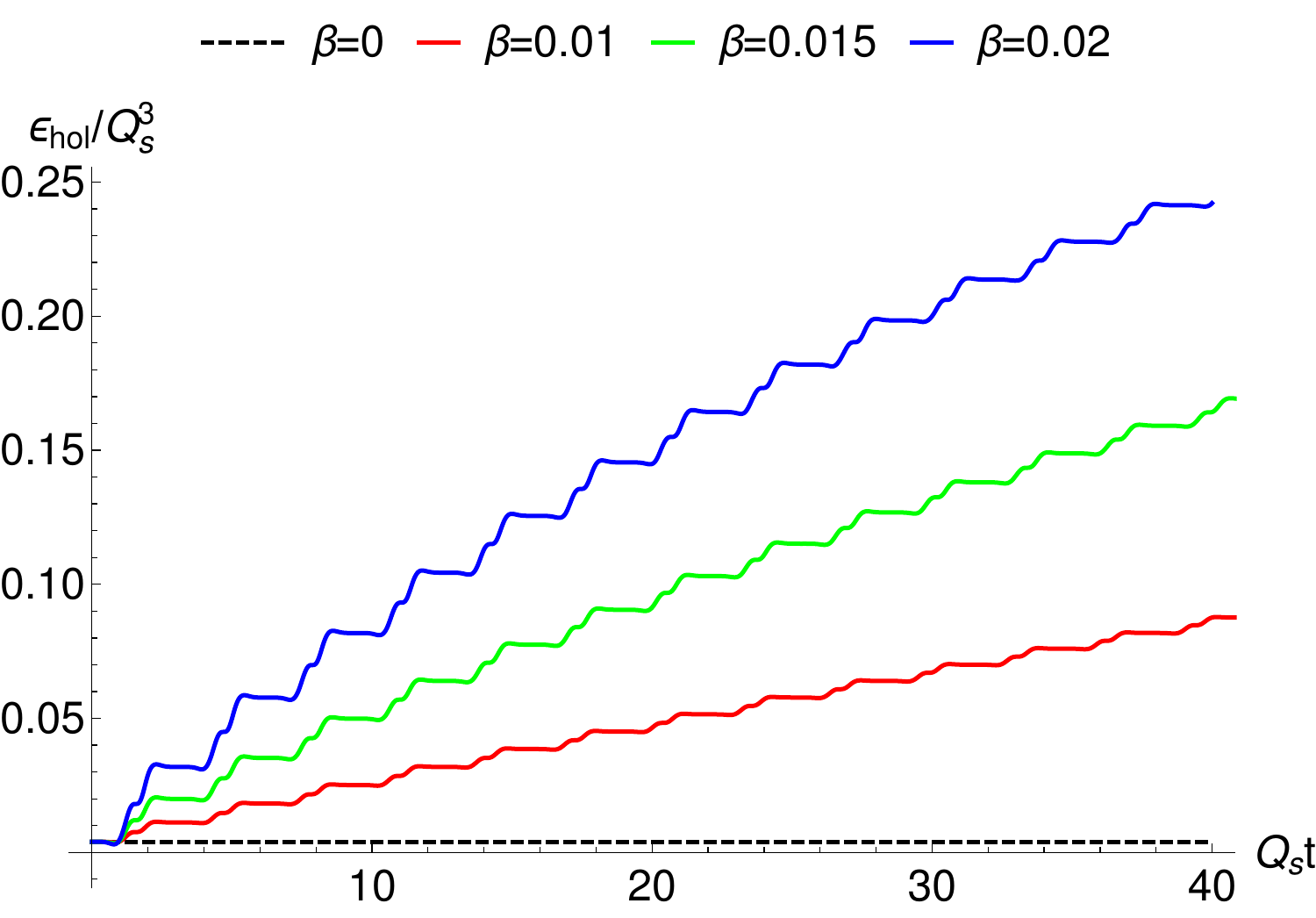}
    \end{minipage}\vfill
       \begin{minipage}{0.55\textwidth}
         \hspace{-0.48cm}
         \includegraphics[width=\textwidth]{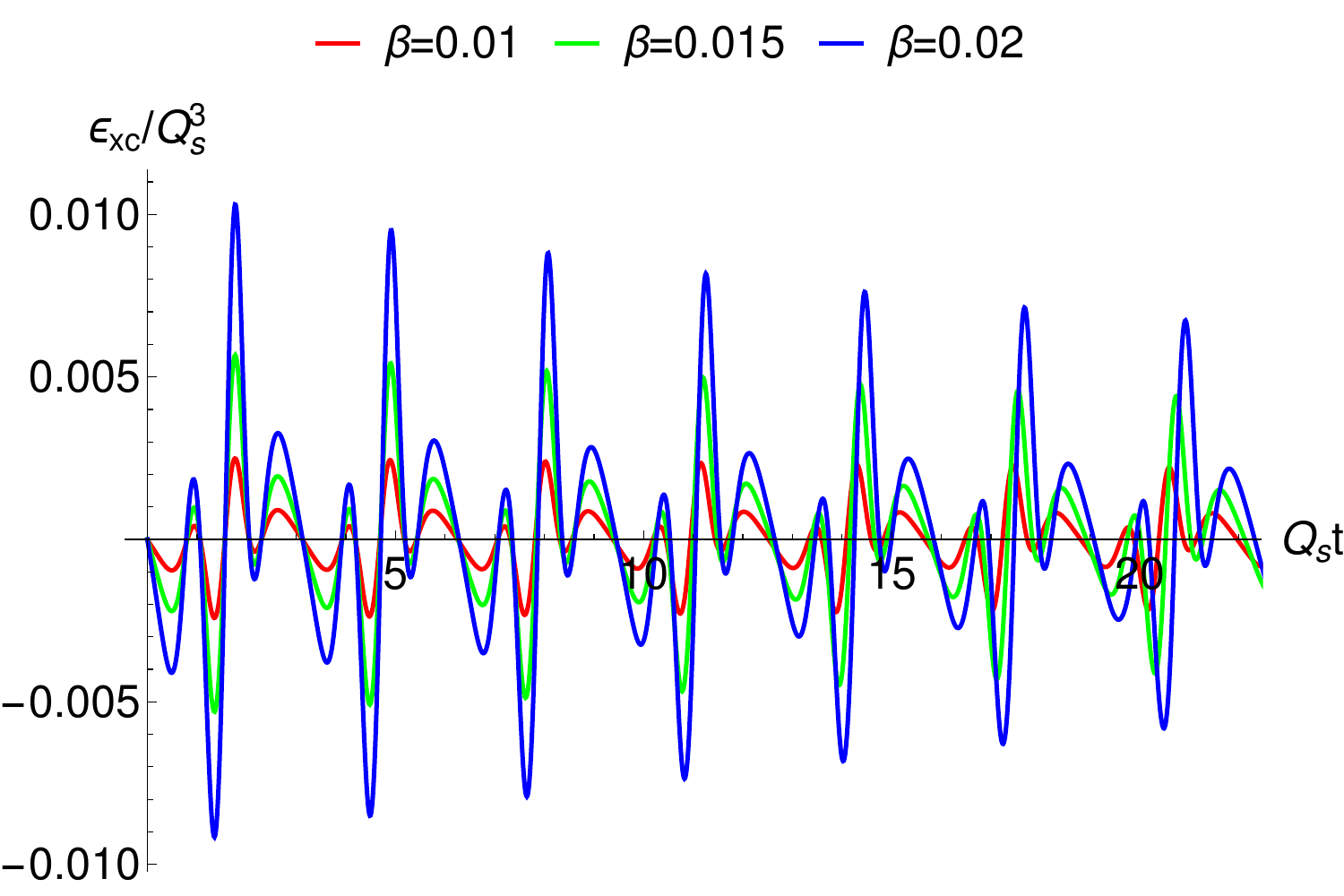}
    \end{minipage}
\caption{Upper: The energy density of the YM sector against time. Middle: The energy density of the holographic sector. Lower: The exchange energy as a function of time.}
\label{energy_densities}
\end{figure}

However, with each iteration the numerical errors in the solution of the respective sub-sectors accumulate, see Appendix \ref{app:1}. We stop the procedure after four iterations in this case, since it provides the optimal trade off between obtaining sufficiently well behaved total energy on the one hand and consistent sub-sectors on the other hand. 

Furthermore, we learn from the plot of $f(t)$ in Fig. \ref{f_vev} that the gauge field decreases in frequency and in amplitude with time. This behavior is more pronounced the higher the coupling $\beta$ is. As a consequence we find that the energy of the Yang-Mills sector on average is decreasing in time, while the energy of the holographic sector is increasing almost monotonically as the first two panels of Fig. \ref{energy_densities} show.

The third panel of Fig. \ref{energy_densities} displays the interaction energy, which like the gauge field oscillates around zero and decays over time with decreasing frequency. Starting from this observation, one might speculate that eventually all energy from the Yang-Mills sector gets transferred to the strongly coupled sector, with decreasing rate. Note that only when the the Yang-Mills sector is empty the source $h(t)$ vanishes and thus the transfer of energy is no longer possible. As long as the source $h(t)$ is varying in time one excites matter fields in the gravitational bulk which fall into the black hole causing its growth.

\begin{figure}[h]
\centering
\begin{minipage}{0.51\textwidth}
        \includegraphics[width=1\textwidth]{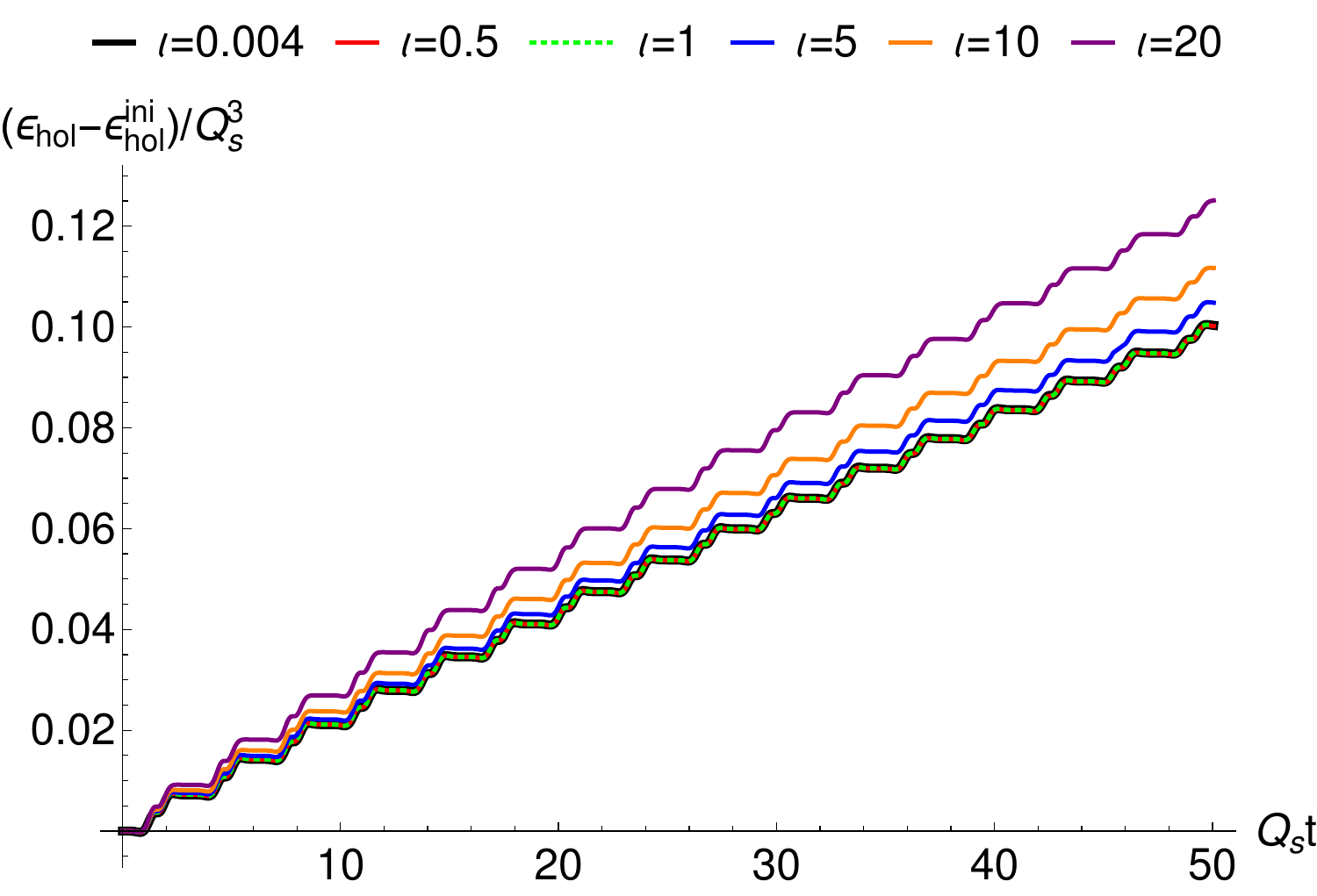}
    \end{minipage}\hfill
    \begin{minipage}{0.47\textwidth}
    \vspace{0.62cm}
    \includegraphics[width=\textwidth]{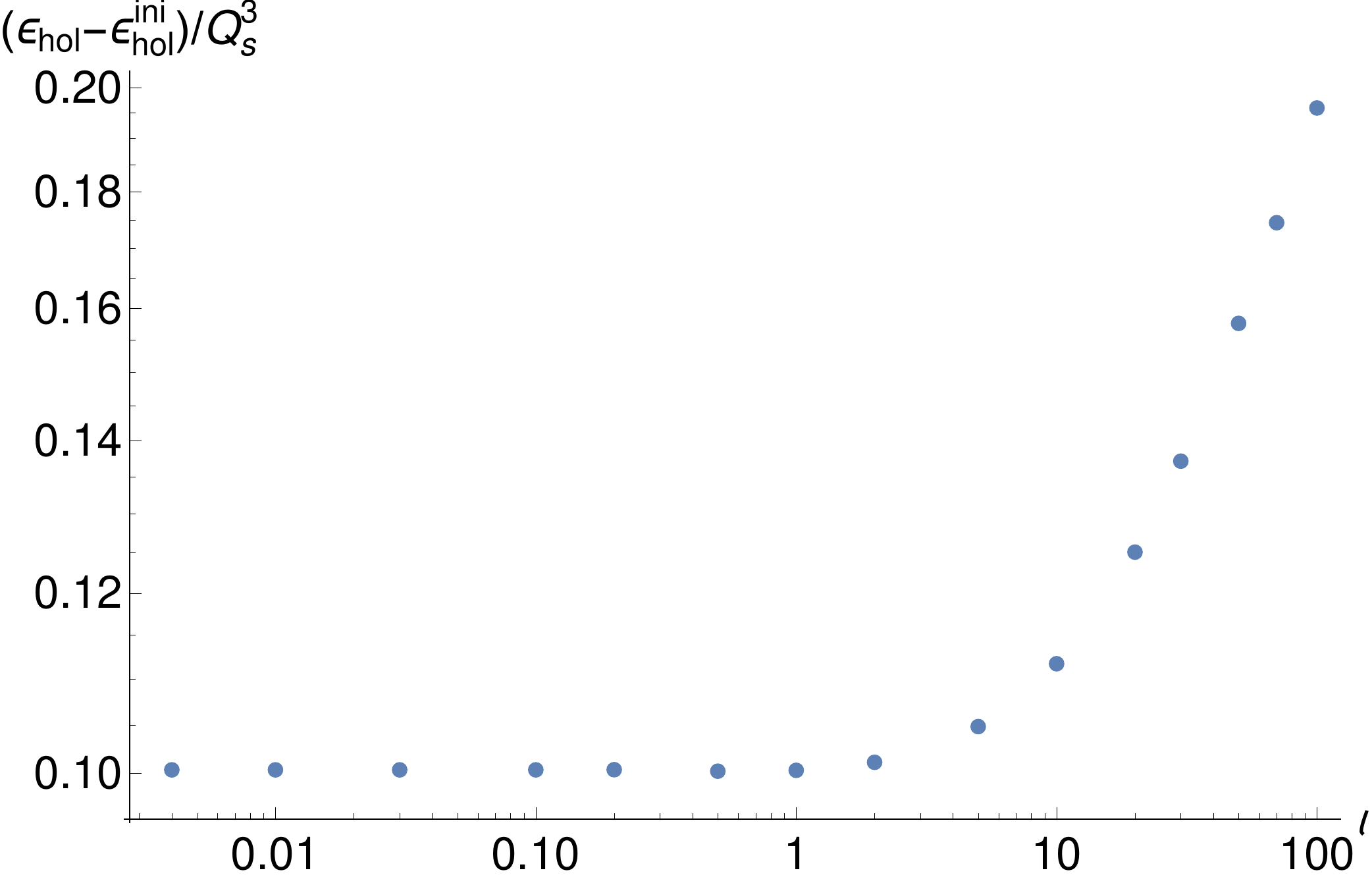}
    \end{minipage}\hfill
\caption{Left: The time evolution of the gain in $\epsilon_\mathrm{hol}$ for different initial conditions $\epsilon_\mathrm{hol}^\mathrm{ini}$ with $\epsilon_\mathrm{YM}^\mathrm{ini}/Q_s^3=1$ and $\beta=0.01$. The curves for $\iota\equiv\epsilon_\mathrm{hol}^\mathrm{ini}/\epsilon_\mathrm{YM}^\mathrm{ini}\le1$ lie on top of each other. Right: 
The gain in $\epsilon_\mathrm{hol}$ at $Q_st=50$ as a function of $\iota$.}
\label{IC}
\end{figure}

The choice for the initial values of the energy in the respective sub-sectors is motivated by the CGC picture of heavy-ion collisions, where the YM sector carries essentially all of the initial energy in the form of highly overoccupied gluons at the saturation scale, but the infrared sector to be described by holography
is initially empty and thus represented by pure AdS spacetime.
Due to numerical issues it is however necessary to start with a small regulator black hole in the gravitational bulk. The two panels in Fig. \ref{IC} compare the gain in the holographic energy for different initial conditions, while the initial Yang-Mills energy is kept fixed. We see that the results are fairly insensitive to the size of this regulator provided that $\iota:=\epsilon_\mathrm{hol}^\mathrm{ini}/\epsilon_\mathrm{YM}^\mathrm{ini}\ll1$ and thus our choice $\iota=0.004$ used for the plots above is a reasonable one.

\begin{figure}
    \centering
    \begin{minipage}{0.5\textwidth}
    \vspace{0.48cm}
        \centering
        \includegraphics[width=0.9\textwidth]{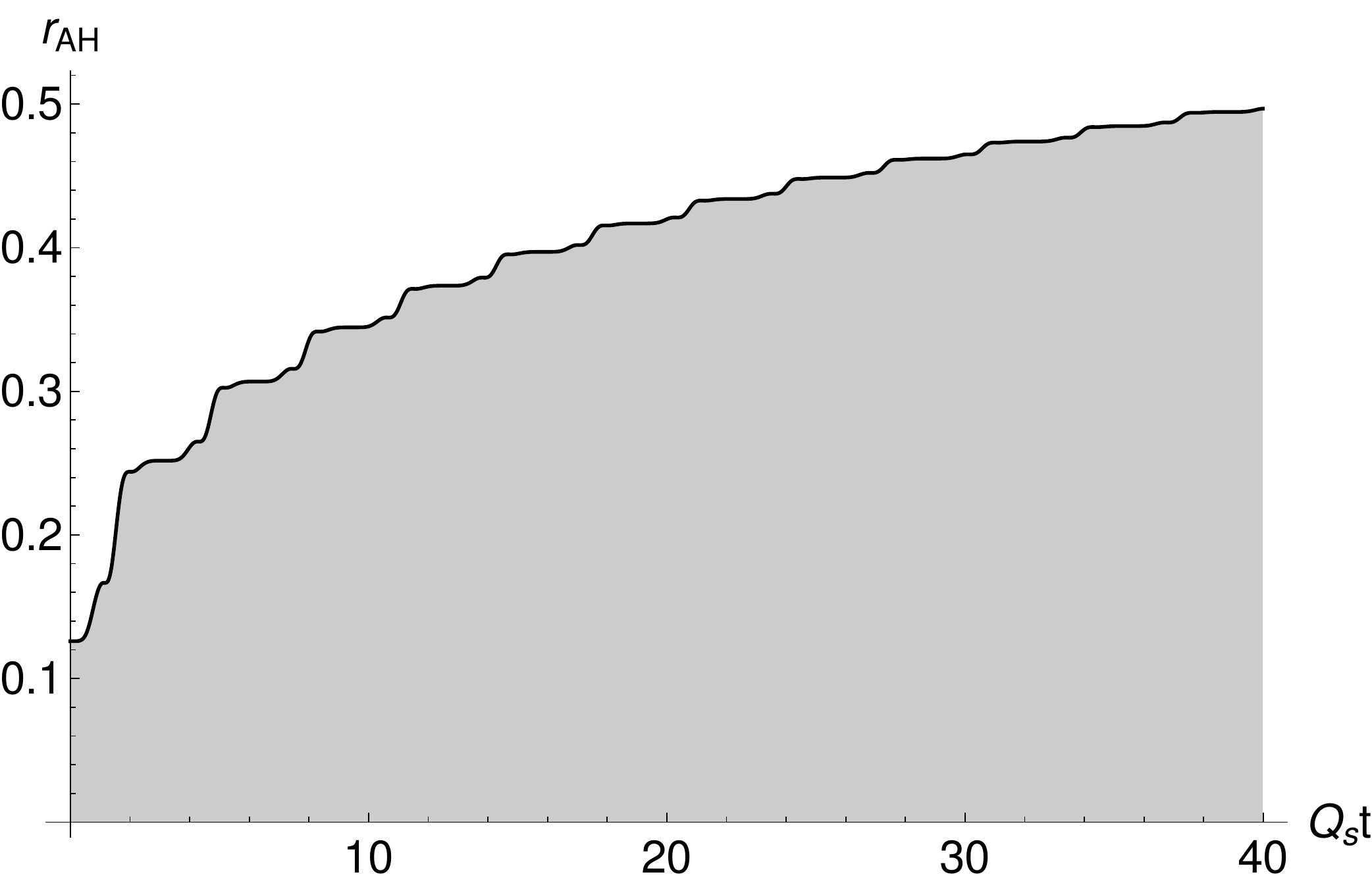}
    \end{minipage}\hfill
    \begin{minipage}{0.5\textwidth}
        \centering
        \includegraphics[width=0.9\textwidth]{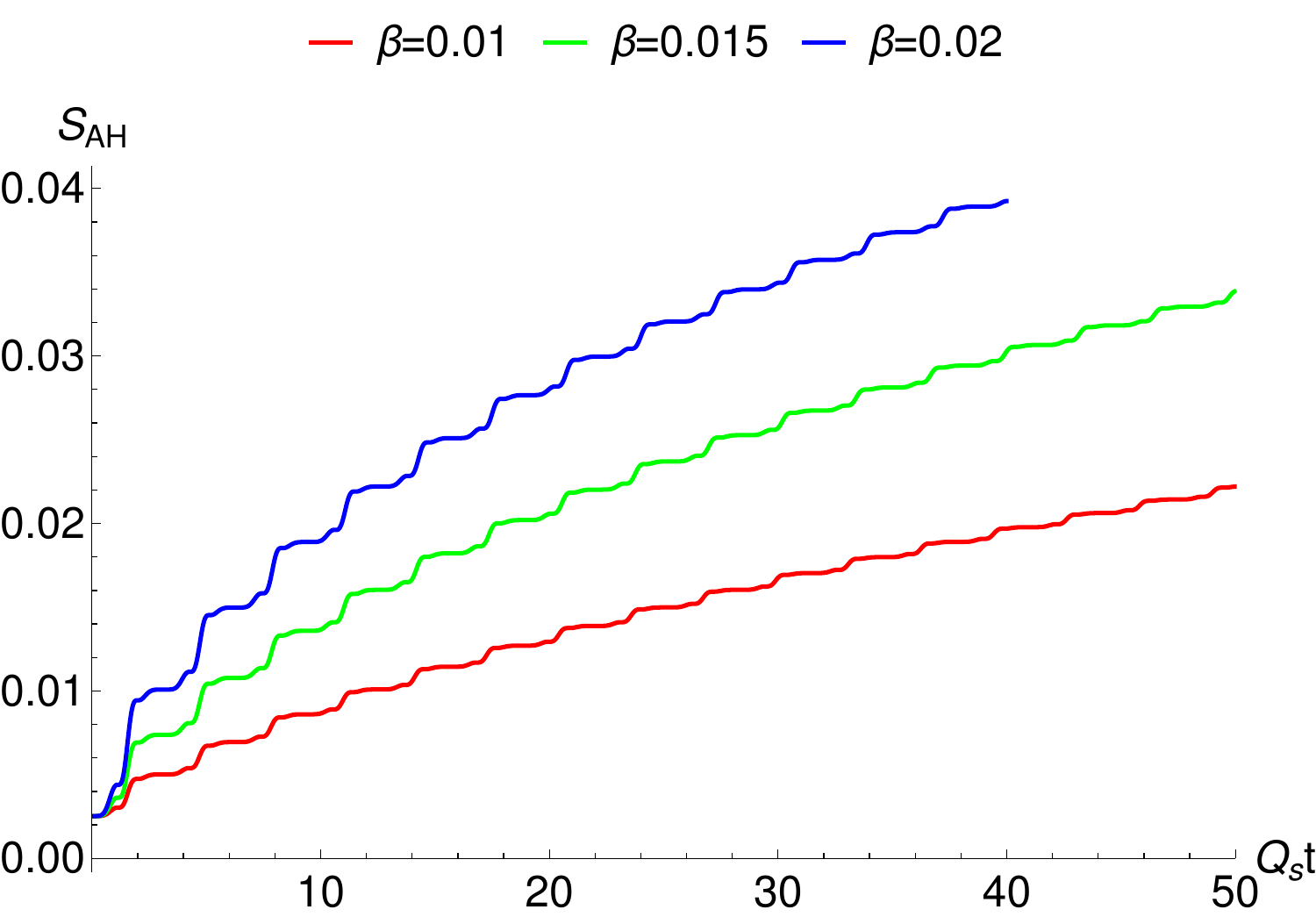}
    \end{minipage}\hfill
\caption{Left: Radial position of the apparent horizon for $\beta=0.02$. The gray region indicates the interior of the black hole. Right:  Entropy in the holographic sector computed from the corresponding areas of the apparent horizons.}
\label{HorizonEntropy}
\end{figure}

In our setup the classical Yang-Mills sector consists only of a single dynamic degree of freedom given by $f(t)$, hence the associated entropy is zero.
However, in the holographic sector the area of the apparent horizon provides a commonly used proxy of entropy, which we use as estimate for the lower bound for the entropy in the combined system. In the left plot of Fig. \ref{HorizonEntropy} we show the radial position of the apparent horizon for the case in which $\beta=0.02$. The right plot of Fig. \ref{HorizonEntropy} shows the entropy associated to the apparent horizon for different values of the coupling $\beta$. We find that the growth of entropy increases with $\beta$. 
Furthermore, we numerically checked that the effective apparent horizon entropy is monotonically increasing with time in all our simulations.

\section{Conclusion and Outlook}
 
In this paper we have extended the toy model used in the first numerical tests \cite{Mukhopadhyay:2015smb} of the semiholographic model for heavy-ion collisions proposed in Ref.~\cite{Iancu:2014ava} such that the glasma equations are coupled to a holographically described infrared sector which permits the formation of a black hole and thus entropy production and thermalization.
The results we discussed present the first successful implementation of a self-consistent numerical AdS/CFT simulation involving a backreacted dynamical boundary source far from equilibrium. As our main result we find that the UV degrees of freedom modelled by a classical Yang-Mills theory lose their energy to the strongly coupled IR degrees of freedom 
over time (see Fig. \ref{energy_densities}), while the total energy is conserved. Motivated by glasma initial conditions one would like to start with all the energy deposited in the Yang-Mills sector and an empty strongly coupled sector. For numerical reasons we have to initialize the gravitational sector with a regulator black hole, but we showed that our results are insensitive to the size of a small regulator. Our numerical findings indicate that eventually the IR degrees of freedom will deplete the energy of the UV degrees of freedom entirely. However, in heavy-ion collisions the glasma picture eventually ceases to be applicable, namely when the occupation numbers of the UV degrees of freedom become of order one or less. At that point a quantum effective kinetic theory description of the UV degrees of freedom should replace the description in terms of classical Yang-Mills fields.

We also want to stress that the model we set up can be applied to contexts other than heavy-ion collisions. In principle one can consistently couple any classical field theory to strongly coupled sources following our scheme.
 
In order to obtain a tractable toy model, we have simplified the semiholographic model by symmetry assumptions and worked in $2+1$ dimension, which made it easier to obtain high numerical accuracy. However, we expect that in $3+1$ dimensions our results will not change qualitatively.

In our future studies we plan to work in $3+1$ dimensions as a next step. To that end we have to improve the numerical stability of our solution procedure. The bottleneck is mainly solving the Yang-Mills equation without introducing too much noise by numerical differentiations. Presently we rely on prebuilt routines of Mathematica when solving the equations, as well as removing some of the noise by filtering techniques.
More importantly, we also want to relax the symmetry assumptions, incorporating anisotropies and spatial inhomogeneities by opening up the spin-2 coupling channel in the semiholographic couplings \cite{Mukhopadhyay:2015smb,Kurkela:2018dku}.

From what we have learned in the present study we expect our model to be very efficient in converting energy in the UV degrees of freedom to a thermal bath represented by a dual black hole, when the sources vary locally in space time.

Note that although the strongly coupled sector is inherently in a quantum regime, our couplings do not couple fluctuations to the glasma. Thus, in order to improve the model from a more conceptual point of view one should implement couplings to quantum fluctuations. This is the subject of work in progress.

\acknowledgments
We would like to thank Daniel Grumiller, Takaaki Ishii, Aleksi Kurkela, and Wilke van der Schee for valuable discussions. We are particularly grateful to Stefan Stricker for contributing to the early stages of this project. A.M.\ acknowledges support from the Ramanujan Fellowship of
DST India and the new faculty initiation grant of IIT Madras.
The research has also been supported by the Austrian Science Fund (FWF) projects P26328-N27, P27182-N27, and W1252-N27.


\appendix

\section{Numerical accuracy of the iterative procedure}\label{app:1}

\begin{figure}[!h]
    \centering
    \begin{minipage}{0.5\textwidth}
        \centering
        \includegraphics[width=0.9\textwidth]{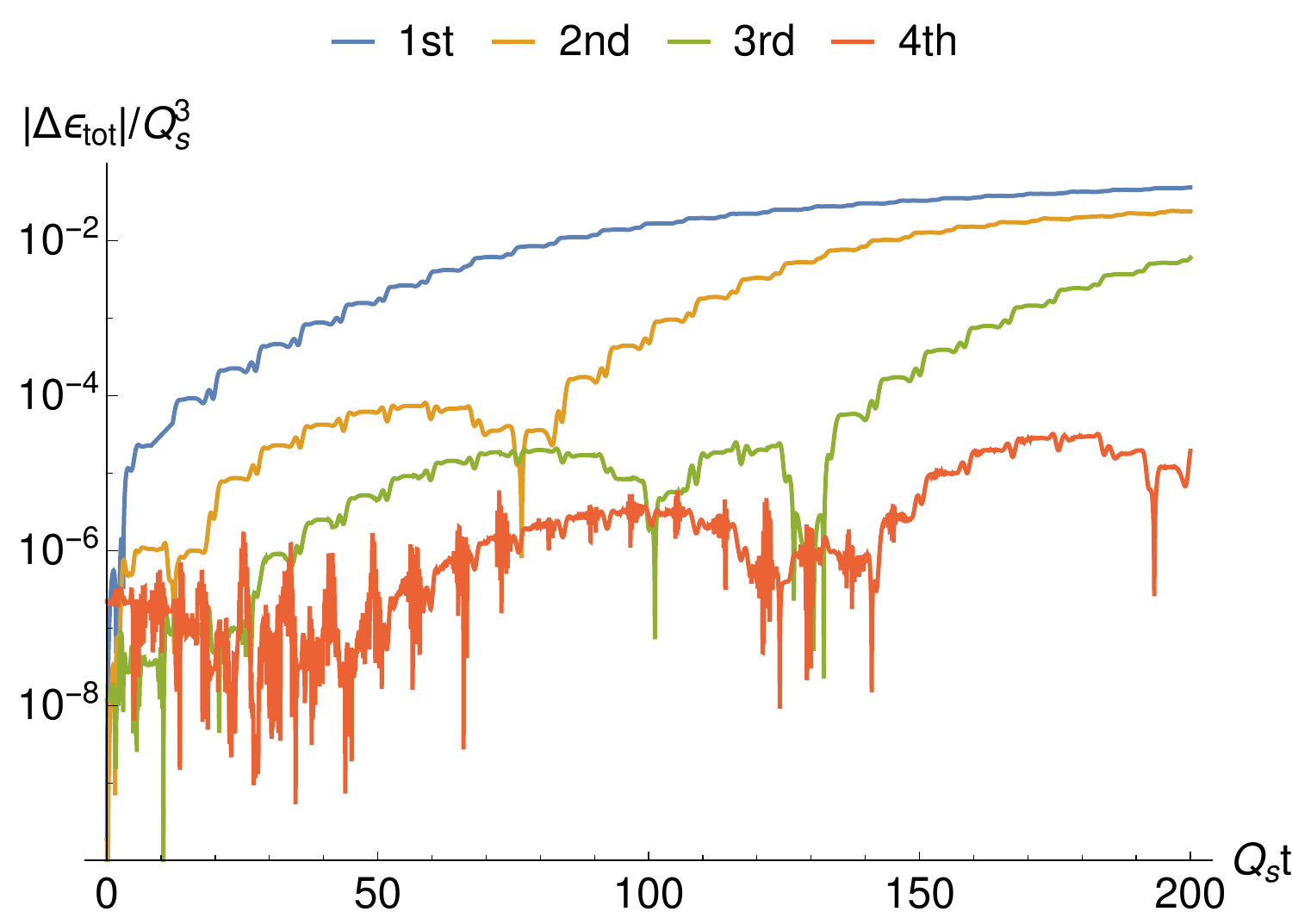}

            \end{minipage}\hfill
    \begin{minipage}{0.5\textwidth}
        \centering
        \includegraphics[width=0.9\textwidth]{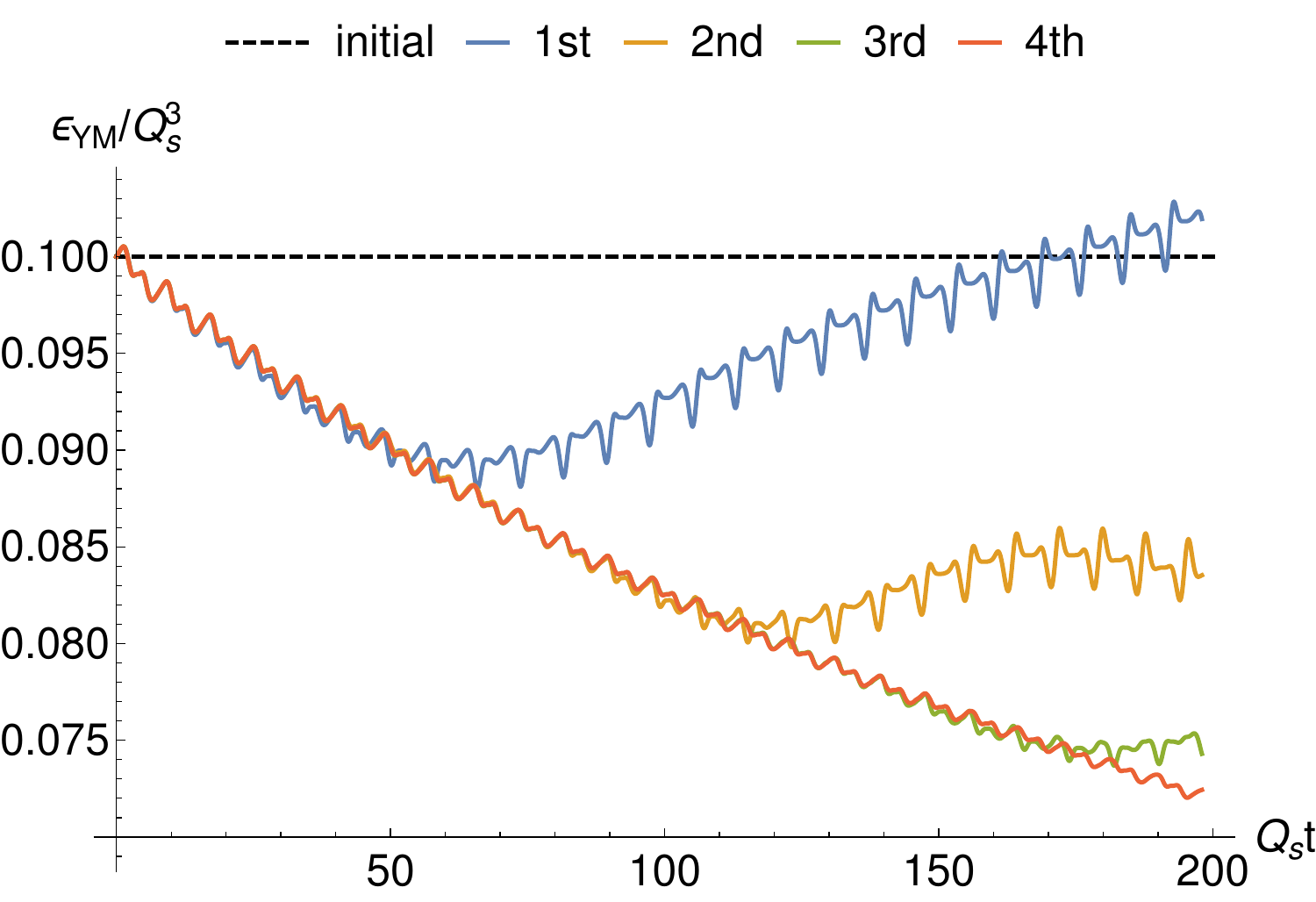}
    \end{minipage}
\caption{Left: Violation of the total energy conservation ($\Delta \epsilon_{tot}(t)=(\epsilon_{YM}^{ini}+\epsilon_{hol}^{ini})-(\epsilon_{YM}(t)+\epsilon_{hol}(t))+\epsilon_{xc}(t)$) as function of time in four subsequent iterations for $\beta=0.2,\coup/\sqrt{Q_s}=0.5,\epsilon_{YM}^{ini}/Q_s^3=0.1$, and $\epsilon_{hol}^{ini}/Q_s^3=1/250$. Right: Time evolution of the energy in the Yang-Mills sector for the same parameters in four subsequent iterations.}
\label{iterations}
\end{figure}

In this appendix we use an illustrative example, characterized by $\beta=0.2,\coup/\sqrt{Q_s}=0.5,\epsilon_{YM}^{ini}/Q_s^3=0.1$, and $\epsilon_{hol}^{ini}/Q_s^3=1/250$, to demonstrate the numerical feasibility of our iterative procedure.
The left plot of Fig. \ref{iterations} shows the violation of total energy conservation $\Delta \epsilon_{tot}(t)=\epsilon_{tot}^{ini}-\epsilon_{tot}(t)$, defined as difference between total initial energy $\epsilon_{tot}^{ini}=\epsilon_{YM}^{ini}+\epsilon_{hol}^{ini}$, and total energy during the time evolution $\epsilon_{tot}(t)=\epsilon_{YM}(t)+\epsilon_{hol}(t)+\epsilon_{xc}(t)$. Note that initially the exchange energy is zero. Typically, after four iterations our numerical scheme arrives at a solution in which $\Delta\epsilon_{tot}(t)/Q_s^3 \approx \mathcal{O}(10^{-5})$ or smaller.

 A characteristic of our algorithm is that at earlier times less iterations are necessary to converge to the true solution than at later times.
This effect is shown in the right panel of Fig. \ref{iterations}, where we plot the Yang-Mills energy in four subsequent iterations.
This behaviour is induced by the way we choose our initial guess, which typically requires less improvement at earlier times, because then it differs not so much in amplitude and phase from the true solution. At later times, when already a significant amount of energy is transferred, guess and true solution can have very different amplitude and phase such that several iterations are necessary to achieve sufficient improvement.
\begin{figure}[!h]
    \centering
    \begin{minipage}{0.5\textwidth}
        \centering
        \includegraphics[width=0.9\textwidth]{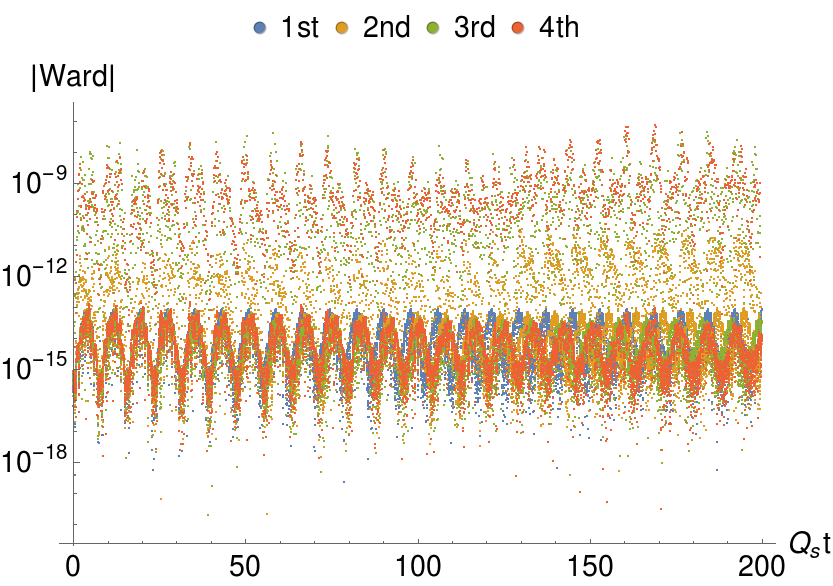}

            \end{minipage}\hfill
    \begin{minipage}{0.5\textwidth}
        \centering
         \includegraphics[width=0.9\textwidth]{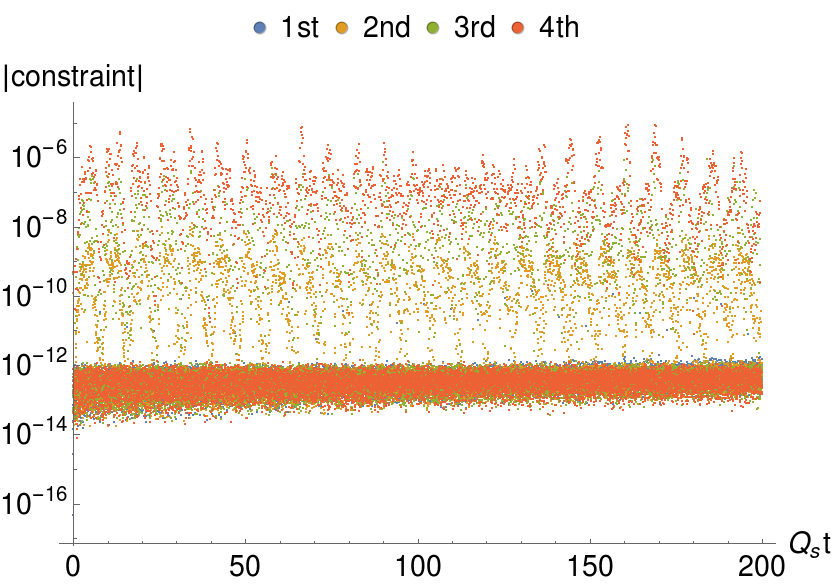}
    \end{minipage}
\caption{Left: Violation of the holographic Ward identity. Right: Constraint in the Einstein equations.}
\label{ward_plot}
\end{figure}

In each of these iterations we monitor the Ward identity \eqref{eq:Ward_hol} and the constraint \eqref{constraint}. In the left plot of Fig. \ref{ward_plot} we see that in most time steps, and in all subsequent iterations, the Ward identity is fulfilled to an accuracy better than $10^{-12}$. In a comparably small number of time steps the accuracy systematically decreases with the number of iterations, but always remains below $10^{-7}$ in this specific example. A similar picture holds for the constraint in the gravity simulation, shown in the right plot of Fig.~\ref{ward_plot}. Also here, in most time steps the absolute value of the maximum violation (in bulk direction) of the constraint \eqref{constraint} remains smaller than $10^{-12}$ in all subsequent iterations, and only for a small number of time steps the error grows with the number of iterations. The origin of this numerical noise, ultimately leading to a break-down of our algorithm, can be traced back to numerical errors introduced when solving the classical Yang-Mills equation \eqref{YM_f}. In particular \eqref{Eq:VEV_hol} shows that derivative of higher order enter the calculation of $\mathcal{H}$, which then in turn complicate the solution of the classical Yang-Mills equation and make the filtering procedure necessary in the first place. Optimizing this part of the simulation is ongoing work.


\bibliographystyle{JHEP}

\bibliography{bibliography}

\end{document}